\documentclass[atoms,article,accept,moreauthors,pdftex,12pt,a4paper]{mdpi}

\lastpage{406}
\doinum{10.3390/atoms3030392}
\pubvolume{3}
\pubyear{2015}
\history{Received: 25 June 2015  / Accepted: 27 August 2015  / Published: 2 September 2015}
\pdfoutput=1

\usepackage[latin1]{inputenc}
\usepackage[T1]{fontenc}
\usepackage{amsmath} 						
\usepackage{amssymb}							
\usepackage{epic}
\usepackage{bm}
\usepackage{soul}
\usepackage{upgreek}

 \theoremstyle{mdpi}
 \newcounter{thm}
 \newcounter{ex}
 \newcounter{re}
 \theoremstyle{mdpidefinition}

\renewcommand{\b}[1]{\mathbf{ #1}}									
\renewcommand{\b}[1]{\mathbf{ #1}}											
\newcommand{\bra}[1]{\langle \, #1 \! \! \mid}							
\newcommand{\ket}[1]{\mid \! \! #1 \,\rangle}							
\newcommand{\up}{\uparrow}														
\newcommand{\down}{\downarrow}												
\newcommand{\rom}[1]{\uppercase\expandafter{\romannumeral #1\relax}} 

\newcommand{\HH}{\mathcal{H}}

\newcommand{\braket}[2]{\langle{#1}|{#2}\rangle}

\newcommand{\realsum}{\displaystyle\sum}

\Title{Probing and Manipulating Fermionic and Bosonic Quantum Gases with Quantum Light}

\Author{Thomas J. Elliott $^{\dagger,}$*, Gabriel Mazzucchi $^{\dagger}$, Wojciech Kozlowski $^{\dagger}$, Santiago F. Caballero-Benitez $^{\dagger}$ and Igor B. Mekhov}

\address[1]{
\scalebox{0.96}{Department of Physics, Clarendon Laboratory, University of Oxford, Parks Road, Oxford OX1 3PU, UK;} E-Mails: {gabriel.mazzucchi@physics.ox.ac.uk (G.M.); wojciech.kozlowski@physics.ox.ac.uk (W.K.); santiago.caballerobenitez@physics.ox.ac.uk (S.F.C.-B.); igor.mekhov@physics.ox.ac.uk (I.B.M.)} \vspace{12pt}}

\contributed{$^\dagger$ These authors contributed equally to this work. }

\corres{E-Mail: thomas.elliott@physics.ox.ac.uk. }

\externaleditor{Academic Editors: Jonathan Goldwin and Duncan O'Dell}

\abstract{We study the atom-light interaction in the fully quantum regime, with the focus on off-resonant light scattering into a cavity from ultracold atoms trapped in an optical lattice. The detection of photons allows the quantum nondemolition (QND) measurement of quantum correlations of the atomic ensemble, distinguishing between different quantum states. We analyse the entanglement between light and matter and show how it can be exploited for realising multimode macroscopic quantum superpositions, such as Schr\"odinger cat states, for both bosons and fermions. We provide examples utilising different measurement schemes and study their robustness to decoherence. Finally, we address the regime where the optical lattice potential is a quantum dynamical variable and is modified by the atomic state, leading to novel quantum phases and significantly altering the phase diagram of the atomic system.}



\begin{document}


\section{Introduction}

Ultracold atomic systems in optical lattices form a powerful tool for studying and simulating the behaviour of quantum degenerate matter. Substantial interest in these systems has arisen partly as a result of their interdisciplinary nature, as they provide exciting possibilities to realise ``toy-model'' Hamiltonians, crucially, with highly tunable parameters, that can describe a wide range of diverse physical systems and processes from many different fields. Such phenomena include quantum phase transitions, strongly correlated materials, high-energy physics \cite{Lewenstein} and biological systems \cite{DieterPRB2013}. Additionally, they provide a strong potential avenue for performing quantum information processing protocols. Light plays a fundamental role in these experiments, as it allows for cooling, trapping and manipulation of the atomic ensemble. Nevertheless, the light field is usually described classically, while its quantum properties are left neglected. However, elevating the treatment of light to the quantum regime leads to many interesting phenomena not present with classical light; pioneering works in the field have already shown that these phenomena include non-destructive quantum measurement \cite{javanainen2003, Mekhov2007b, Roscilde2009, Rogers2014, RuostekoskiAF2014, Kozlowski2015}, self-organisation and other novel phase behaviour \cite{PolzikNatPh2007, LarsonPRL2008, MeystrePRA2009, Morigi, Ivanov2014, Caballero2015, Sachdeva2015}, as well as the possibility to engineer quantum states and dynamics, particularly through the measurement backaction effect \cite{ruostekoski1998, Bhattacherjee2009, mekhovPRA2009, mekhovLP2011, Pedersen2014, Lee2014, Mazzucchi2015, Elliott2015}.

In this article, we exploit the opportunities offered by the quantum nature of light for use in probing quantum correlations and realising macroscopic quantum superposition states. To this end, we build upon earlier work on bosons \cite{Mekhov2012}, expand studies of quantum light as a quantum non-demolition measurement device and as a method of state preparation to the fermionic case and make use of some novel possibilities that arise as a result of their spinful nature. We present a homodyne detection scheme, which may be used to engineer states, including Schr\"odinger cat states. Moreover, we also study some of the properties of the light-matter entanglement that forms a key ingredient in making such effects and protocols possible. Further, we show how the quantum coupling between light and matter modifies the atomic trapping potential, leading to novel quantum phases and allowing for the implementation of tunable long-range interactions not achievable with \emph{s}-wave Feshbach resonances.

In analogy (yet contrastingly) to cavity QED experiments \cite{HarocheBook}, where the quantum light states in a cavity were probed and projected by measuring atoms, here, the roles are reversed, and we probe properties of matter fields with light. One such advantage this approach yields is in the ease of scalability; the number of atoms can be tuned from a few to thousands by changing the number of illuminated sites, while scaling cavities is a challenge \cite{HarocheBook}. Thus, this work, alongside recent experiments involving Bose--Einstein condensates (BECs) in optical cavities \cite{EsslingerNat2010, HemmerichScience2012, ZimmermannPRL2014}, where such predictions can be tested, helps to bridge the gap between quantum optics and quantum gases, approaching the fully-quantum regime of many-body light-matter interactions \cite{Mekhov2012,ritsch2013}.

\begin{figure}[H]
\centering
\includegraphics[width=.8\textwidth]{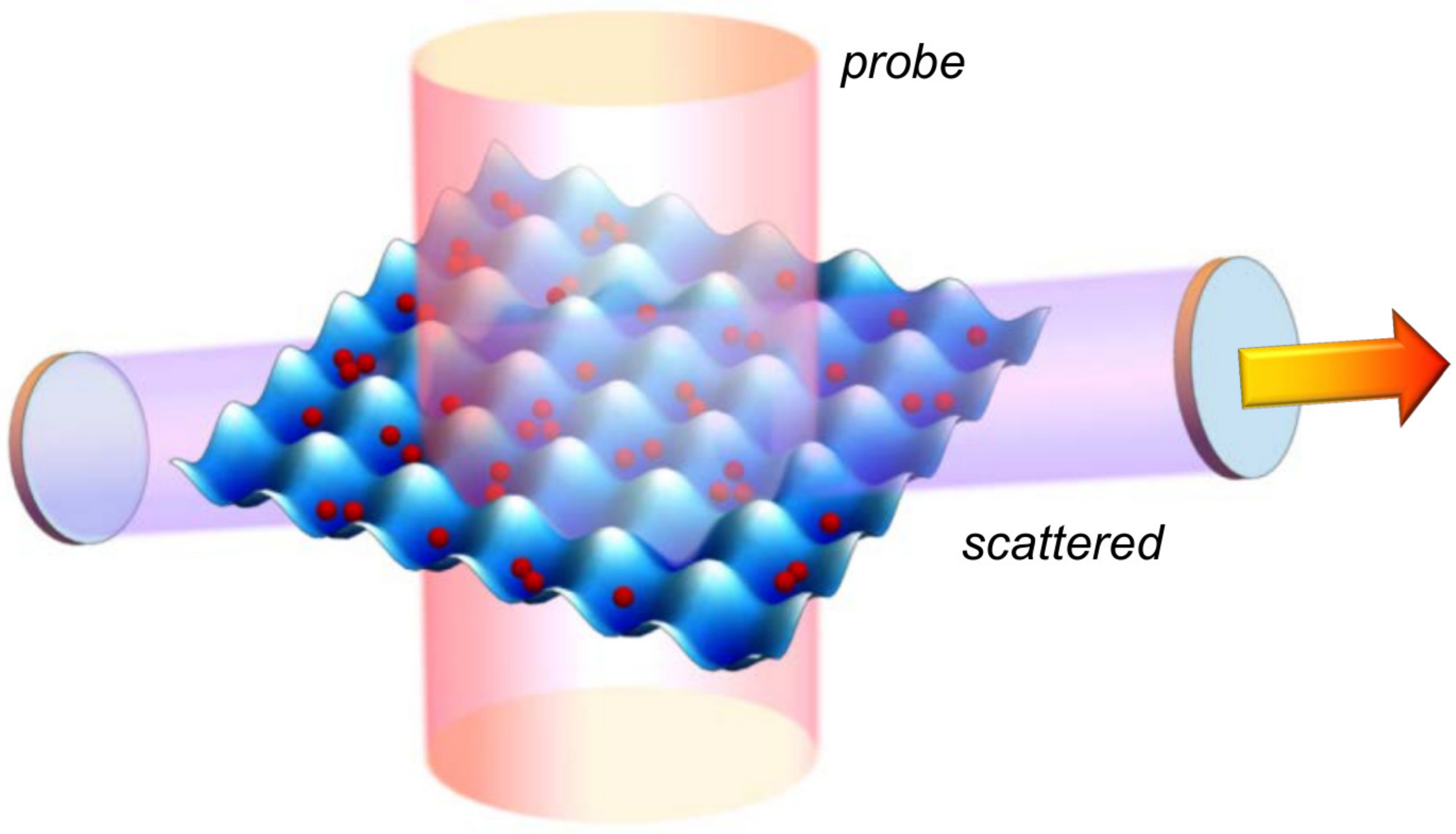}
\caption{Setup. Atoms in an optical lattice are illuminated by a probe beam ({red}), and the scattered light ({blue}) is selected and enhanced by an optical cavity.}
\label{fig:setup}
\end{figure}

We focus on a system of ultracold atoms trapped in an optical lattice with $M$ sites and coupled to quantum light. Specifically, the atoms are illuminated with an off-resonant coherent probe beam with amplitude $\upalpha_0$ and frequency $\upomega_p$; the scattered light at a particular angle can be selected and enhanced by an optical cavity with decay rate $\upkappa$ and mode frequency $\upomega_c$ (Figure \ref{fig:setup}). The Hamiltonian of the full light-matter system is $\HH=\HH^A+\HH^{L}+\HH^{AL}$. $\HH^A$ describes the atomic dynamics in an optical lattice~\cite{Lewenstein}. For bosons, this is the Bose--Hubbard Hamiltonian:

\vspace{-12pt}
\begin{equation}
\HH^A_b=-t_0\sum_{\langle i,j\rangle}(\hat b^\dagger_i\hat b^{\phantom{\dagger}}_j+h.c.)+\frac{U}{2}\sum_i\hat n_i(\hat n_i -1)
\end{equation}

and for fermions, the Hubbard model:
\begin{equation}
\HH^A_f=-t_0\sum_{\upsigma\in\{\up,\down\}}\sum_{\langle i,j\rangle}(\hat f^\dagger_{i\upsigma}\hat f^{\phantom{\dagger}}_{j\upsigma}+h.c.)+U\sum_i\hat n_{i\up}\hat n_{i\down}
\end{equation}
where $\hat b_i$ and $\hat f_{i,\upsigma}$ are the bosonic and fermionic annihilation operators at site $i$, respectively, $t_0$ is the nearest-neighbour hopping amplitude, $U$ is the on-site atom-atom interaction energy, the bosonic (fermionic) number operator $\hat{n}_i=\hat{b}_i^\dagger \hat{b}_i$ ($\hat{n}_{i\upsigma}=\hat{f}_{i\upsigma}^\dagger \hat{f}_{i\upsigma}$) and the sum $\langle i,j\rangle$ is taken over neighbouring pairs of sites. The light field is described by $\HH^{L}=\hbar\upomega_c\hat a_1^\dagger \hat a_1$, where $\hat{a}_1$ is the cavity mode annihilation~operator.

In the bosonic case, the light-matter interaction is given by~\cite{Mekhov2012}:
\begin{equation}
\HH^{AL}_b=\hbar \Omega_{11} \hat a_1 ^\dagger \hat a_1 \hat F_{11} +\hbar \Omega_{10} \left(\upalpha_0^*\hat a_1\hat F_{10}^\dagger+\upalpha_0\hat a_1^\dagger\hat F_{10}\right)
\end{equation}
where the atoms couple to the cavity mode via the coefficient $\Omega_{lm}= g_l g_m /\Delta_a$, $g_l$ is the light-matter coupling constant for mode $l$, $\Delta_a$ is the detuning between light and the atomic resonance, $\hat F_{lm}=\hat D_{lm}+\hat B_{lm}$ and:
 \begin{align}
\hat D_{lm}=\sum_{i=1}^M J_{ii}^{lm} \hat n_i, \qquad \hat B_{lm}=\sum_{\langle i,j\rangle }^M J_{ij}^{lm} \hat b_i^\dagger \hat b_j
\end{align}

The coefficients $J_{ij}^{lm}$ characterise the coupling between light and matter at each lattice site and are defined as the convolution between the light mode functions $u_l(\b{r})$ and the localised atomic Wannier functions $w(\b{r})$, \textit{i.e}.,
\begin{align}
J_{ij}^{lm}=\int \! w(\b{r}-\b{r}_i)u^*_l(\b{r})u_m(\b{r})w(\b{r}-\b{r}_j)\, d \b{r}
\end{align}

Note that the operator $\hat D_{lm}$ describes the scattering from on-site density, while $\hat B_{lm}$ that from small intersite densities~\cite{Kozlowski2015}. For well-localised atoms, $J_{ij}^{10}\approx u^*_1(\b{r}_i)u_0(\b{r}_j) \updelta_{ij}$, and the contribution to the light scattering due to $\hat B_{lm}$ can often be neglected.

In the fermionic case, we introduce the light polarisation as an additional degree of freedom for the light field, which allows for probing of different spin species. In particular, we use linearly-polarised light $a_{1x}$ and $a_{1y}$, which couple differently to the two spin densities $\hat{n}_{i\uparrow}$ and $\hat{n}_{i\downarrow}$ because of selection rules that constrain the allowed transitions between different hyperfine states of the atoms~\cite{Greiner2005,Lee2012a,Meineke2012,Sanner2010,Sanner2011,Sanner2012}. In contrast to the (spinless) bosonic case, this allows us to probe different linear combinations via the operators $\hat{D}_{lm,x} = \sum_i J_{ii}^{lm} \hat{\uprho}_i$ or $\hat{D}_{lm,y} = \sum_i J_{ii}^{lm} \hat{m}_i$ where $\hat{\uprho}_i = \hat{n}_{i\uparrow}+\hat{n}_{i\downarrow}$ (density) and $\hat{m}_i = \hat{n}_{i\uparrow}-\hat{n}_{i\downarrow}$ (magnetisation). In the fermionic case, we also consider attractive particle-particle interactions. Additionally, due to the differing particle statistics and Pauli exclusion, fermions exhibit different dynamics and phase behaviour to their boson counterparts.

\section{Quantum Nondemolition Probing of Many-Body States}

The atom-light coupling in the Hamiltonian $\HH^{ml}$ introduces correlations between light and matter. As~a consequence, it is possible to characterise the quantum state of the atoms by performing measurements of the light observables. Specifically, probing the light amplitude $\hat a_1$ or photon number $\hat a_1 ^\dagger \hat a_1$ describes a quantum nondemolition (QND) measurement of the observables related to $\hat D_{lm}$ \cite{Brune1992}. We focus on the regime where the atomic dynamics (\textit{i.e}.,~tunnelling) is much slower than the light scattering: the tunnelling amplitude $t_0$ and the interaction $U$ determine the quantum properties of the atomic state, but such properties do not change during the measurement time. The stationary solution of the Heisenberg equations for the scattered light operator $\hat a_1$ (neglecting the small dispersive frequency shift $\Omega_{11}\hat D_{11}$) is given by:
\begin{align}
\hat a_1=\frac{\Omega_{10} \upalpha_0}{\Delta_p + i\upkappa}\hat D_{10} \equiv C \hat D
\end{align}
where $\Delta_p=\upomega_p - \upomega_c$ is the probe-cavity detuning. Analogously to classical physics, the light amplitude $\langle\hat a_1\rangle$ depends on the mean density at each lattice site and describes the scattering of the probe beam $\upalpha_0$ from a periodic grating of atoms. Its angular distribution is determined by the position of the atoms: $\langle\hat a_1\rangle$ reaches maximum value if they scatter light with the same phase, while it vanishes if the contributions from each lattice site interfere destructively. Moreover, quantum density-density correlations of the atomic state are imprinted on the expectation value of the photon number operator:
\begin{align}
\langle\hat a_1^\dagger \hat a_1 \rangle = |C|^2 \langle \hat{D}^\dagger \hat{D} \rangle = |C|^2 \sum_{i,j} J_{ii}^* J_{jj} \langle \hat{n}_i \hat{n}_j \rangle
\end{align}

Higher moments of $\hat a_1$ probe higher-order correlation functions (e.g.,~the fluctuations of the photon number depend on the four-point density expectation values). In order to emphasise the difference between classical and quantum scattering contributions, we focus on two-point correlation functions and~define:
\begin{align}\label{Rdef}
R= \langle \hat{D}^\dagger \hat{D} \rangle - |\langle \hat{D}\rangle|^2 = \sum_{i,j} J_{ii}^* J_{jj} \left(\langle \hat{n}_i \hat{n}_j \rangle - \langle \hat{n}_i \rangle \langle \hat{n}_j \rangle \right)
\end{align}

\begin{figure}[H]
\centering
\includegraphics[width=0.72\textwidth]{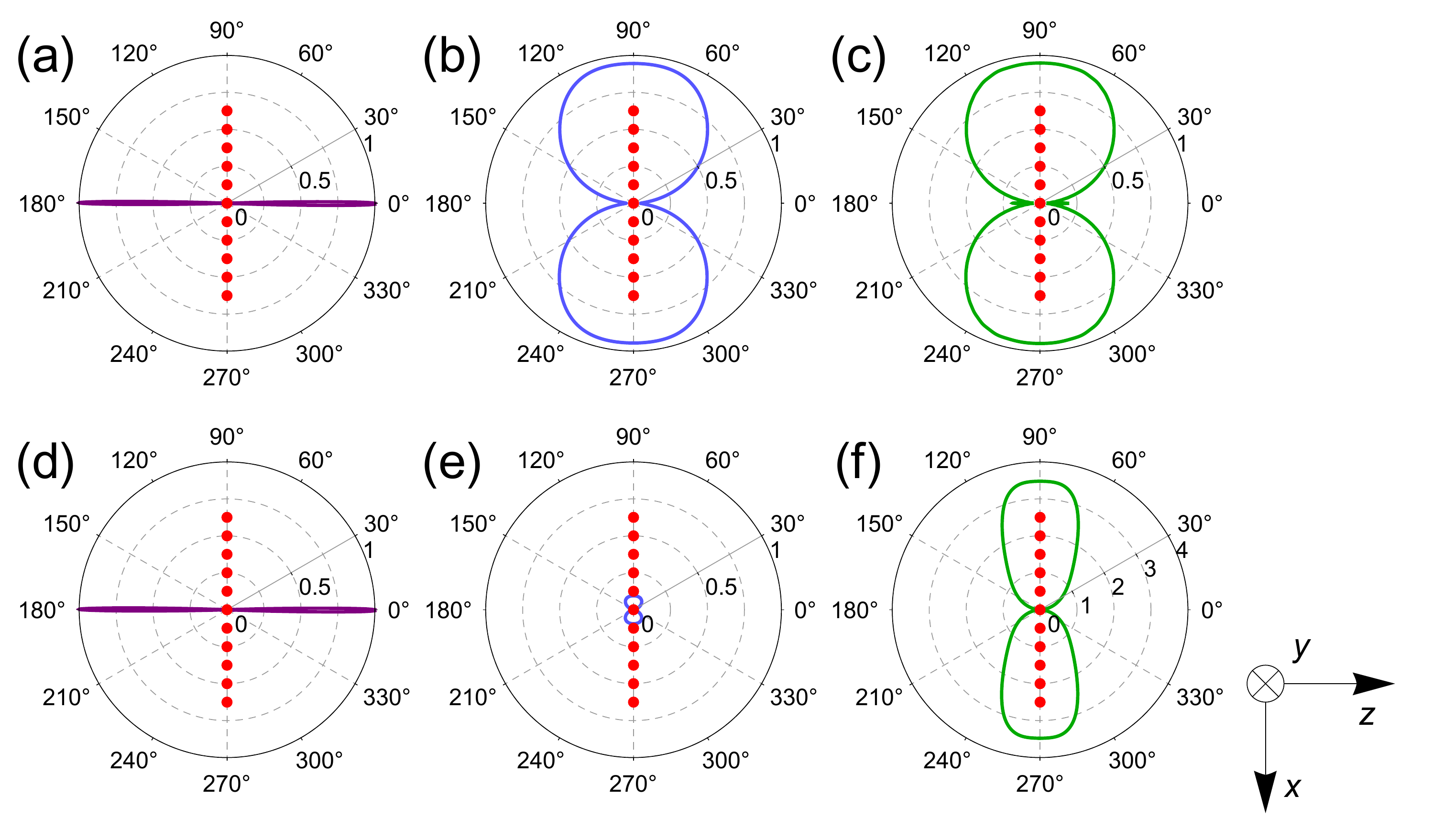}
\caption{Classical diffraction patterns (normalised to $N^2$) and quantum additions (normalised to $N$) for $N$ fermions in a one-dimensional optical lattice (45 sites, half filling) in the non-interacting regime (first row) and for $U/t_0=10$ (second row). The lattice extends into the vertical direction, and the probe is along the horizontal, \textit{i.e.},~the angle marked with $0^{\circ}$ indicates the light intensity in the forward direction. The classical diffraction patterns (\textbf{a},\textbf{d}) do not depend on the interaction while the quantum additions $R_y$ (\textbf{b},\textbf{e}) and $R_x$ (\textbf{c},\textbf{f}) distinguish between different ground states. Note that the scale in ({f}) is different from ({a}--{e}): the fluctuations in the atom number increase as the attraction between the atoms favours doubly-occupied sites. Red circles indicate the orientation of the lattice.}
\label{fig:addition}
\end{figure}

This quantity depends on the scattering angle via the coefficients $J_{ii}$ and is directly related to the fluctuations of the atom number. Therefore, quantum light scattering distinguishes between different quantum states, such as the superfluid and Mott insulator states \cite{Mekhov2012}. As mentioned in the previous section, we extend the bosonic model \cite{Mekhov2012,MekhovLP09, mekhovLP2013} to also describe spin-1/2 fermionic atoms and define the quantum additions $R_x$ and $R_y$, related to the fluctuations in density and magnetisation, respectively.
If~the mode functions of the probe and scattered light are travelling waves with wave-vectors $\b{k_\mathrm{in}}$ and $\b{k_\mathrm{out}}$, respectively, the coefficients $J_{jj}$ are proportional to $\exp{\left[ i (\b{k_\mathrm{in}}-\b{k_\mathrm{out}}) \cdot \b{r}_j \right]}$, and the quantum addition $R$ is proportional to a structure factor, \textit{i.e}.,~ the Fourier transform of the density-density correlations. Figure \ref{fig:addition} compares the classical diffraction pattern and quantum additions $R_x$ and $R_y$ for fermions in a one-dimensional optical lattice at half filling. The scattering patterns were calculated for the ground state, obtained via imaginary time evolution using the TNT
 library \cite{TNT}. In particular, we describe the system using the Hubbard model and focus on the ground state of the system in two different regimes: non-interacting ($U/t_0=0$) and strongly-attractive interactions ($U/t_0=10$). Note that in both cases, the local magnetisation of the system is zero ($\langle \hat{m}_i \rangle=0$), and as a consequence, the classical diffraction pattern for linear-$y$ polarised light vanishes, \textit{i.e}., $ \langle \hat a_{1y} \rangle=0$. Nevertheless, the quantum addition $R_y$ is non-zero and depends on the quantum state of the system. In particular, we find that $R_y$ decreases with increasing values of $U/t_0$, as the attraction between the atoms favours doubly-occupied sites, \textit{i.e}.,~the formation of pairs of fermions with opposite spin, which suppress the fluctuations in the magnetisation. Furthermore, the classical diffraction pattern due to the atomic density does not depend on $U/t_0$, as the local density is independent of the interaction for a translationally-invariant optical lattice ($\langle \hat{\uprho}_i \rangle=1$). However, the presence of doubly-occupied sites increases the fluctuations in the atom number, leading to a stronger signal for $R_x$. In general, the scheme we propose allows direct probing of the phase diagram for the superfluid-Mott insulator-Bose glass phases in the one-dimensional Bose--Hubbard model \cite{Kozlowski2015} and the detection of different quantum states of atomic systems such as dimers \cite{Caballero2015}, density waves and magnetic order.

\section{Measurement Backaction and State Preparation}

Due to the entanglement between light and atoms, the detection of a photon affects the dynamics of the atomic state. This is one of the most important manifestations of quantum mechanics, the so-called measurement back action. In order to study this effect, we focus on the dynamics of the quantum state conditioned to the measurement process in a single experimental realisation and describe it using the quantum trajectory formalism.
We consider fermions in a one-dimensional optical lattice, inside an optical cavity with decay $\upkappa_y$, such that only photons linearly polarised along the $y$ axis are allowed to escape it, and thus, the measurement scheme probes the magnetisation of the atomic sample. We illuminate $K$ lattice sites at equal intensity with a coherent beam, and we detect the $y$-polarised scattered light in the diffraction maximum, effectively probing the magnetisation of the illumined sites as $a_{1y}=C(\hat{N}_{K\up}-\hat{N}_{K\down})=C \hat{M}_K$, where $N_K$ and $M_K$ are the total atomic occupation number and magnetisation of all illuminated sites. If the $i$-th photon is detected at time $t_i$, the state of the system changes instantaneously: the cavity annihilation operator $\hat {a}_{1y}$ is applied to it as $\ket{\Psi_c(t_i)}\rightarrow \hat {a}_{1y}\ket{\Psi_c(t_i)}$ (and the state is subsequently normalised). Moreover, the evolution of the system between two consecutive photo counts is determined by the non-Hermitian Hamiltonian:
\begin{align}
\HH_{\mathrm{eff}}=\HH^{L}+\HH^{AL}-i \upkappa_y \hat a_{1y}^\dagger \hat{a}_{1y}
\end{align}

The photo count events and the evolution due to $\HH^{\mathrm{eff}}$ have opposite contributions to the expectation value of $\hat M_K$, as the former tends to increase the magnetisation of the illuminated area, while the latter tends to decrease it. The full dynamics of the system is therefore conditioned by the photodetections occurring at times $t_1, t_2$\ldots..., which are determined stochastically. The initial wavefunction of the light-matter system can be written:
\begin{align}
\ket{\Psi(0)} = \sum_{\b{k},\b{q}} c_{\b{q k}}^{(0)}\ket{\b{q} \up} \ket{\b{k} \down} \ket{\upalpha_{\b{q k}R}(0)}\ket{\upalpha_{\b{q k}L}(0)}
\end{align}
where $\ket{\upalpha_{\b{q k}p}(0)}$ are the coherent states of the light field with polarisation $p$ (in the circular polarisation basis) and $\ket{\b{q} \upsigma}$ are the Fock states for fermions of species $\upsigma$. Since we are neglecting atomic tunnelling, the Hamiltonian $\HH^{\mathrm{eff}}$ does not mix the different atomic Fock states, and it is possible to write an analytical expression for the conditional evolution of $\ket{\Psi_c(t)}$. Specifically, the coherent states $\ket{\upalpha_{\b{q k}p}(0)}$ acquire a phase factor $\exp [\Phi_{\b{qk}}(t)]$, which depends only on the atomic configuration $\b{qk}$ \cite{Mekhov2012,mekhovPRA2009,mekhovLP2010, mekhovLP2011}: this phase reaches a steady state value after $t\gg1/\upkappa_y$ for each quantum trajectory, allowing us to write the conditional wavefunction of the system after $m$ photodetection events as:
\begin{align}\label{cond}
\ket{\Psi_c(m,t)} \propto \sum_{\b{k},\b{q}} c_{\b{q k}}^{(0)} \, (\upalpha_{\b{kq}R} - \upalpha_{\b{qk}L})^{m} \, \mathrm{e}^{\Phi_\b{qk}(t)} \ket{\b{q} \up} \ket{\b{k} \down}\ket{\upalpha_{\b{qk}R}(t)}\ket{\upalpha_{\b{qk}L}(t)}
\end{align}

Due to the steady state in all of the light amplitudes $\upalpha_{\b{qk}p}$, this expression does not depend on the specific detection times $t_i$. Interestingly, the state of the system is not factorisable into a product of matter and light states as the two remain entangled. Moreover, the cavity introduces an effective interaction between the two spin species, since different atomic configurations with different spins are coupled as a result of the photodetection. Focussing on the properties of the atomic system, we compute the probability distribution of having $z_\upsigma$ fermions with spin $\upsigma$ in the illuminated area of the optical lattice at time $t$. This can be obtained summing the absolute values of the coefficients of all of the configurations $\b{q}^\prime\b{k}^\prime$ with the same $z_\upsigma$ and introducing the initial probability distribution $P^{(0)}(z_\upsigma)=\sum_{\b{q}^\prime,\b{k}^\prime} |c_{\b{q}^\prime \b{k}^\prime}^{(0)}|^2$.
Hence, the conditional probability distribution is:
\begin{align}
P_c(t,z_\up,z_\down)=\frac{1}{\mathcal{N}} |z_\up -z_\down|^{2m} \exp{\left( -\tau| z_\up - z_\down|^2 \right)} \, P^{(0)}(z_\up) P^{(0)}(z_\down)
\end{align}
where $\mathcal{N}$ is a normalisation constant and $\tau= 2 |C|^2 \upkappa_y t$. Rewriting this expression in terms of the magnetisation of the system, we have:
\begin{align}
P_c(t,M_K)=\frac{1}{\mathcal{N}} |M_K|^{2m}\exp{\left( -\tau M_K^2 \right)} \, P^{(0)}(M_K)
\end{align}

\begin{figure}[H]
\centering
\includegraphics[width=12cm]{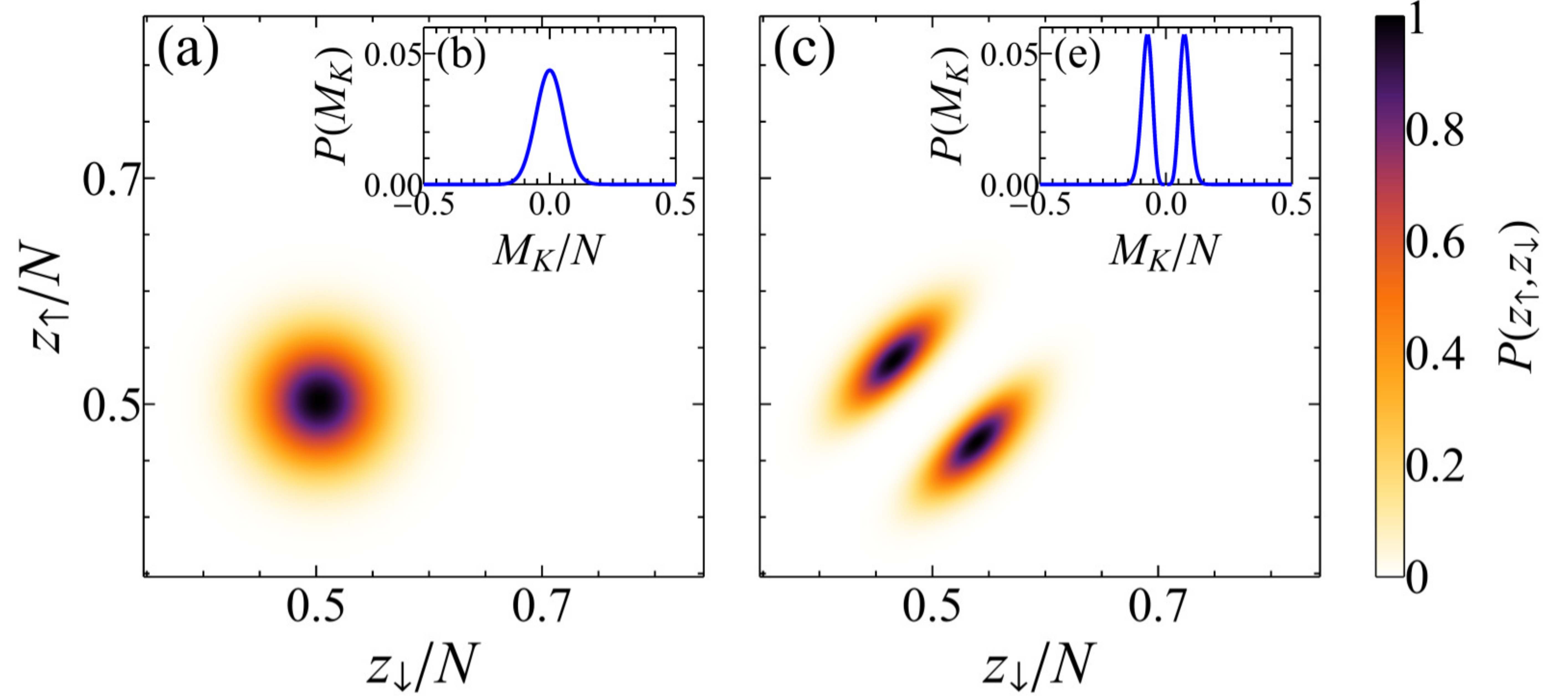}
\caption{Probability distribution $P(z_\up,z_\down)$ without measurement (\textbf{a}) and with measurement (\textbf{c}) for a fermionic system. The insets show the probability distribution of the magnetization $P(M_K)$ for the initial state (\textbf{b}) and after photons are detected (\textbf{e}). Measurement backaction creates a Schr\"odinger cat state.}
\label{fig:cat}
\end{figure}

The relation between $m$ and $t$ follows a stochastic process, which defines a single quantum trajectory. The measurement process strongly modifies $P^{(0)}(M_K)$: when the first photon is detected, this probability distribution becomes bimodal, and all of the configurations with $M_K=0$ are forbidden (Figure \ref{fig:cat}). Moreover, as time progresses, $P(t,M_K)$ shrinks to two narrow peaks around the values $M_{1,2}=\pm \sqrt{m/\tau}$ with decreasing width (note that the stochastic nature of this expression arises through $m$). Thus, the final state of the system has a well-defined absolute magnetisation as the measurement process projects the atomic state to a superposition of two states with magnetisations $M_1$ and $-M_1$: a Schr\"odinger cat state. This multicomponent structure emerges as a consequence of the degeneracy of the photon number operator $\hat a_{1y}^\dagger \hat a_{1y}$ and can be exploited to realise non-trivial quantum superpositions by detecting photons at different angles \cite{Mekhov2012} or by engineering the coefficients $J_{ii}$ using different light mode functions \cite{Elliott2015,Mazzucchi2015}.

\section{Light-Matter Entanglement}

The light-matter entanglement is a key feature of the system; indeed, it is this entanglement that is responsible for many of the effects observed when one goes beyond semiclassical treatments. It is thus interesting to consider the properties of this entanglement and how it is affected by the measurement backaction. As we deal with pure states, we can characterise the (bipartite) entanglement properties with the entanglement entropy \cite{amico2008}, defined as the Von Neumann entropy of the reduced density matrix of one of the subsystems:
\begin{equation}
E(\ket{\Psi}_{AB})=S(\hat\uprho_A)=-\mathrm{Tr}(\hat\uprho_A\log\hat\uprho_A)
\end{equation}

If the reduced density matrix is diagonal, this is equivalent to the Shannon entropy $H(\hat\uprho)=-\sum_i P_i\log P_i$, where $P_i$ are the diagonal elements of $\hat\uprho$. When the number of atoms is large, we can approximate the distribution of subsystem occupation numbers as continuous, such that we may invoke the central limit theorem and treat them as Gaussian functions \cite{boas2006}. For such functions, the Shannon entropy is given by:
\begin{equation}
\label{eqcontEoE}
H(P)=\frac{1}{2}\log(2\pi e \upsigma^2)
\end{equation}
where $\upsigma^2$ is the variance of the distribution.

We first consider the entanglement present in the steady state for bosons in a lossless cavity. In~this case, when the lattice is illuminated by a coherent probe, the state of the system can be written $\ket{\Psi(t\to\infty)}=\sum_{\b{q}}c_{\b{q}}\ket{\b{q}}\ket{\upalpha_{\b{q}}}$, and hence, the reduced state of the light as:
\begin{equation}
\uprho_L=\realsum_{\b{q}}|c_{\b{q}}|^2\ket{\upalpha_{\b{q}}}\bra{\upalpha_{\b{q}}}
\end{equation}

Thus, the entanglement entropy can be obtained directly from the Shannon entropy of the distribution $P(z)=\sum_{\b{q}|\upalpha_{\b{q}}=\upalpha_z}|c_{\b{q}}|^2$, where $z$ represents all states $\b{q}$ corresponding to the same $\upalpha_{\b{q}}$ (\textit{i.e}., $z$ is the light measurement eigenvalue), and we have assumed that light states corresponding to different $z$ are approximately orthogonal. We justify this by noting the overlap of two coherent states is given by $|\braket{\upalpha_1}{\upalpha_2}|^2=e^{-|\upalpha_1-\upalpha_2|^2}$ and, hence, is exponentially suppressed by the difference in the $\upalpha$, which can be tuned by, for example, increasing the pump power.

We investigate two paradigmatic examples where there is one incoming probe beam and one scattered beam. In the first, we illuminate a region of the lattice at the diffraction maximum ($z=N_K$) and
the second at the diffraction minimum ($z=N_K^{\mathrm{even}}-N_K^{\mathrm{odd}}$). When the matter is initially in a superfluid state, for the diffraction maximum, we have that $z$ obeys a Poisson distribution $P(z)=e^{-\langle N\rangle}\langle N\rangle^z/z!$, where $\langle N\rangle$ is the expected number of atoms in the illuminated region, while for the diffraction minimum, it obeys a Skellam distribution $P(z)=e^{-\langle N\rangle}I_z(2\langle N\rangle)$, where $I_z$ is a modified Bessel function of the first kind, and we have assumed that the expected number of atoms on odd and even sites is equal. Since both of these distributions have the same variance $\upsigma^2=\langle N\rangle$, from Equation \eqref{eqcontEoE}, we find that for both cases, the light-matter entanglement entropy is:
\begin{equation}
E=\frac{1}{2}\log(2\pi e\langle N\rangle)
\end{equation}

When cavity loss is taken into account, the $c_{\b{q}}$ and, hence, $P(z)$ evolve, according to $P(z,t)=|\upalpha_z^m|^2e^{-2|\upalpha_z|^2\upkappa t}P(z,0)/\mathcal{N}$ \cite{Mekhov2012}. This squeezes the initial distribution, and hence, the light-matter entanglement will decrease over time and eventually vanish, as the distribution of $z$ converges to a single value; hence, the system enters a product of the light and matter states.

However, a peculiarity arises in the diffraction minimum case if we consider only the leaked photon count rate to have been detected and the phase not measured. In doing so, only the magnitude of $\upalpha_z$ is constrained, and there is no distinction between $z$ and $-z$. Thence, at long times, the state of the system becomes a cat state \cite{Mekhov2012}:
\begin{align}\label{cat1}
\ket{\Psi(\tau,m\to\infty)}& \frac{1}{\sqrt{2}}(\ket{z}\ket{\upalpha_z}+(-1)^m\ket{-z}\ket{\upalpha_{-z}})
\end{align}
possessing a light-matter entanglement entropy of one independent of the number of atoms, $z$, and the light field strength.

\section{Homodyne Detection}

\begin{figure}[H]
 \centering
 \includegraphics[width=13cm]{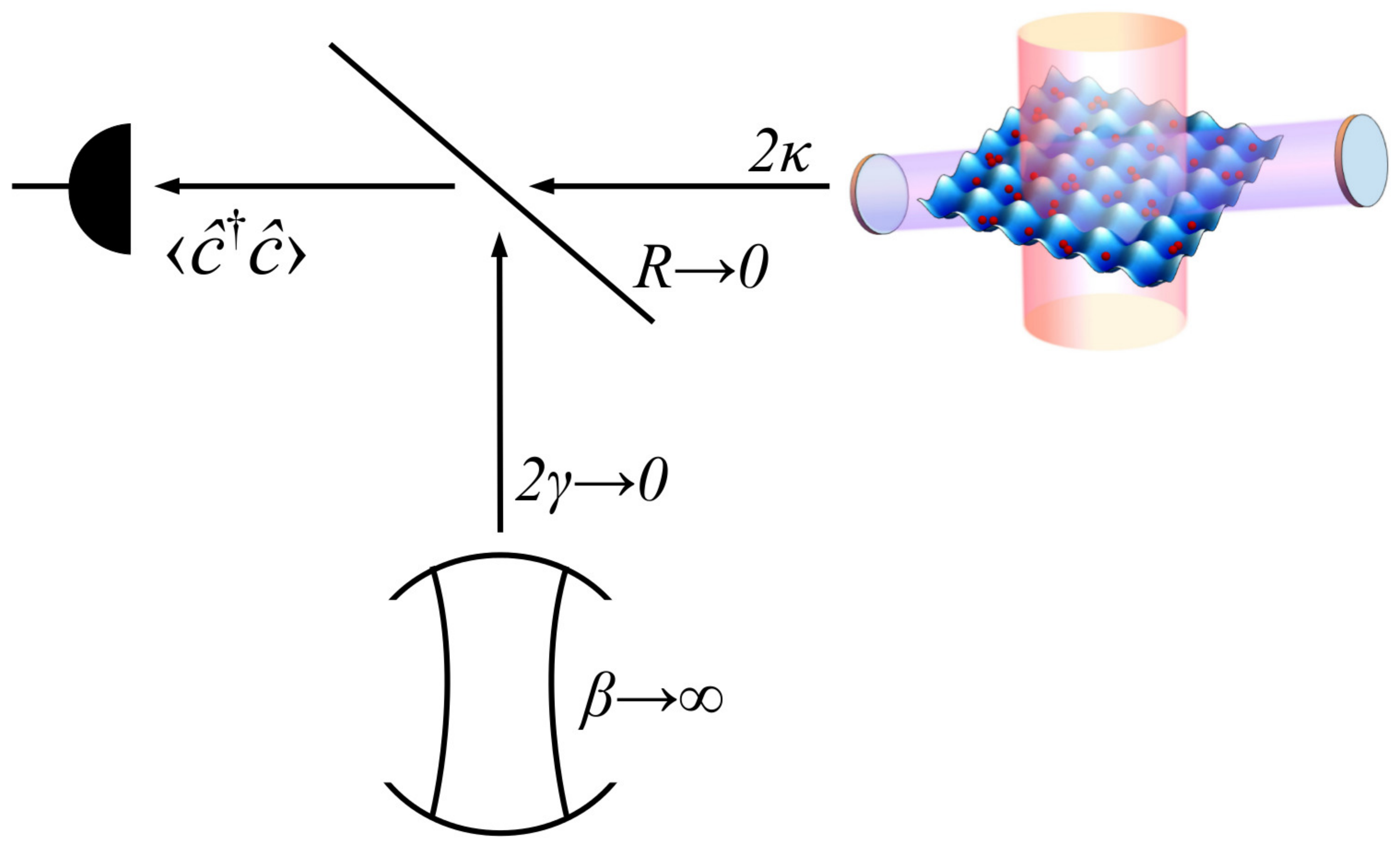}
 \caption{The homodyne detection scheme as analysed in \cite{Carmichael}. The pumped cavity contains an atomic gas in an optical lattice potential. The second cavity radiates a coherent local oscillator field $|\upbeta \rangle$.}
 \label{fig:Homodyne}
\end{figure}

The measurement scheme introduced in the first section of this article is extremely flexible and can be easily extended. For example, we consider a setup that includes a local oscillator cavity and
another cavity containing the atomic system, as illustrated in Figure~\ref{fig:Homodyne}, where $\upkappa$ and $\upgamma$ are cavity decay constants for the atomic system and local oscillator, respectively, and
$\upbeta$ is the local oscillator coherent amplitude. We combine the two light modes with a beam splitter with reflectivity $R$, and in order for all of the atomic cavity photons to reach the detector in our system, we let $R \rightarrow 0$. However, this also requires that the local oscillator amplitude, $\upbeta \rightarrow \infty$ and $\upgamma\rightarrow 0$, such that the local oscillator flux at the detector, $\mathcal{F}= R^2\upgamma|\upbeta|^2$, remains constant \cite{Carmichael}. Assuming that the atomic dynamics is much slower than the light scattering, it can be shown that the evolution of the light-matter system is described by the non-Hermitian~Hamiltonian:
\begin{equation}
 \HH_\mathrm{eff} = \HH^{l} + \HH^{ml} - i \hbar \upkappa \hat a_1 ^\dagger \hat a_1 - i \hbar e^{-i \theta} e^{i\upomega_p t} \sqrt{2 \upkappa \mathcal{F}} \hat a_1
\end{equation}
where $\uptheta$ is the local oscillator phase at the detector. In this case, the jump operator that is applied to the quantum state at each photodetection is \cite{Carmichael}:
\begin{equation}
 \hat c=\sqrt{\mathcal{F}} e^{i \uptheta} e^{-i\upomega_p t} +\sqrt{2 \upkappa} \hat a_1
\end{equation}

The state of the system at time $t$ after $m$ photodetections is:
\begin{equation}
 | \Psi_c(m,t) \rangle \propto c^0_{z_+}\left[1 + e^{i \updelta \phi} z_+ \right]^m e^{i\varphi t} |z_+ \rangle | \upalpha_{z_+} \rangle + c^0_{z_-}\left[1 + e^{i \updelta \phi} z_- \right]^m e^{-i\varphi t} |z_- \rangle | \upalpha_{z_-} \rangle
\end{equation}
where $\varphi$ is a constant, which can be derived from the Hamiltonian,
\begin{align}
z_\pm &= \sqrt{\frac{\mathcal{F}}{2 \upkappa |C|^2}} \zeta_\pm \\
\zeta_\pm &= \pm \sqrt{ \frac{m/t}{\mathcal{F}} - \sin^2(\updelta \phi)} - \cos(\updelta \phi)
\end{align}
where $\updelta \phi = \phi_C - \uptheta$ is the phase difference between $\upalpha_0$ and the local oscillator, $C = |C|e^{i \phi_C}$ and $z$ are the unique eigenvalues of the $\hat{D}$ operator, and as such, they are a suitable label of the measurement~eigenstates.

The homodyne measurement scheme allows us the freedom to choose the phase difference between the local oscillator and the photons emitted from the atomic cavity system, $\updelta \phi = \phi_C - \uptheta$, by adjusting the local oscillator phase. There are two interesting special cases depending on the value of this phase difference. Firstly, we can set the difference to an integer multiple of $2\pi$, $\updelta \phi = 2 \pi n$. In this case, the final state is given by:
\begin{equation}\label{fragile}
 | \Psi_c(m,t) \rangle \propto c^0_{z_+} e^{i \varphi t} |z_+ \rangle | \upalpha_{z_+} \rangle + c^0_{z_-} (-1)^m e^{-i\varphi t} |z_- \rangle | \upalpha_{z_-} \rangle
\end{equation}
with $z_\pm = \sqrt{1 / 2 \upkappa |C|^2} ( \pm \sqrt{ m/t } - \sqrt{\mathcal{F}})$, which is a cat state, albeit still entangled with the light. The light and matter can then be disentangled, though, by switching off the probe and detecting all leaked photons, leaving the cavity in the vacuum mode \cite{mekhovPRA2009}. However, each photo count flips the phase difference between the two components by $\pi$, making this state particularly fragile with respect to decoherence: with each missed photo count, the state loses its purity and becomes a mixed state, losing any atomic entanglement it possessed. Another notable case is if the two phases are offset by $\pi/2$, $\updelta \phi = \pi/2 + 2 \pi n$. In this case, \mbox{$z \equiv z_+ = -z_- = \sqrt{1 / 2 \upkappa |C|^2} \sqrt{ m/t - \mathcal{F}}$}, and the state evolution conditioned to the measurement leads to:
\begin{equation}\label{robust} \small
 | \Psi_c(m,t) \rangle \propto c^0_{z}\left[\sqrt{\mathcal{F}} + i \sqrt{ m/t -\mathcal{F}}\right]^m e^{i \varphi t} |z \rangle | \upalpha_{z} \rangle + c^0_{-z}\left[\sqrt{\mathcal{F}} - i \sqrt{ m/t - \mathcal{F}}\right]^m e^{-i \varphi t} |- z \rangle |- \upalpha_{z} \rangle
\end{equation}

Importantly, $m/t - \mathcal{F}$ is small, and each photo count imparts only a small change in phase between the two components. The state Equation (\ref{robust}) is therefore more robust than Equations (\ref{cat1}) and (\ref{fragile}), as it does not suffer much from the decoherence problem associated with a missed photo count, and this scheme may prove a more useful method to prepare cat states \cite{mekhovPRA2009}.

\section{Quantum Optical Lattices}

The quantum properties of light also induce a modification to the optical lattice potential experienced by the atoms. This leads to an effective quantum potential that changes the potential energy surface: the optical lattice becomes a quantum dynamical variable that depends on the atomic state. The effective Hamiltonian in the steady state of light is \cite{EPJD08,Morigi,Caballero2015}:
\begin{eqnarray}\label{qpot}
 \HH^A_\mathrm{eff}=\HH^A+g_{\mathrm{eff}}\hat D^\dagger \hat D
 \label{effmodel}
\end{eqnarray}
where $g_{\mathrm{eff}}= \hbar \Delta_p |C|^2$. The cavity introduces an effective long-range interaction, which allows the system to support novel emergent orders, even in a single-mode cavity~\cite{Caballero2015}.
Therefore, the properties of the ground state of $\HH^m_\mathrm{eff}$ do not depend only on the local processes (tunnelling and on-site interaction) present in the usual (Bose) Hubbard Hamiltonian, but also on the detuning $\Delta_p$ via the $\hat D$-operator. This drastically changes the effective steady state of the system. It is instructive to analyse the simplest measurement scheme, where atoms scatter light homogeneously into a single-mode cavity. In this case, there is no spatial variation in the density couplings $J_{ii}=J_D$, and thus, there is no density wave instability. Without quantum light and in the absence of disorder, the system supports superfluid and Mott-insulating phases \cite{Lewenstein}. However, the quantum potential will distort the phase diagram, and it is able to shift the superfluid-Mott insulator transition conditional on the light amplitude.

\begin{figure}[H]
\centering
\includegraphics[width=0.7\textwidth]{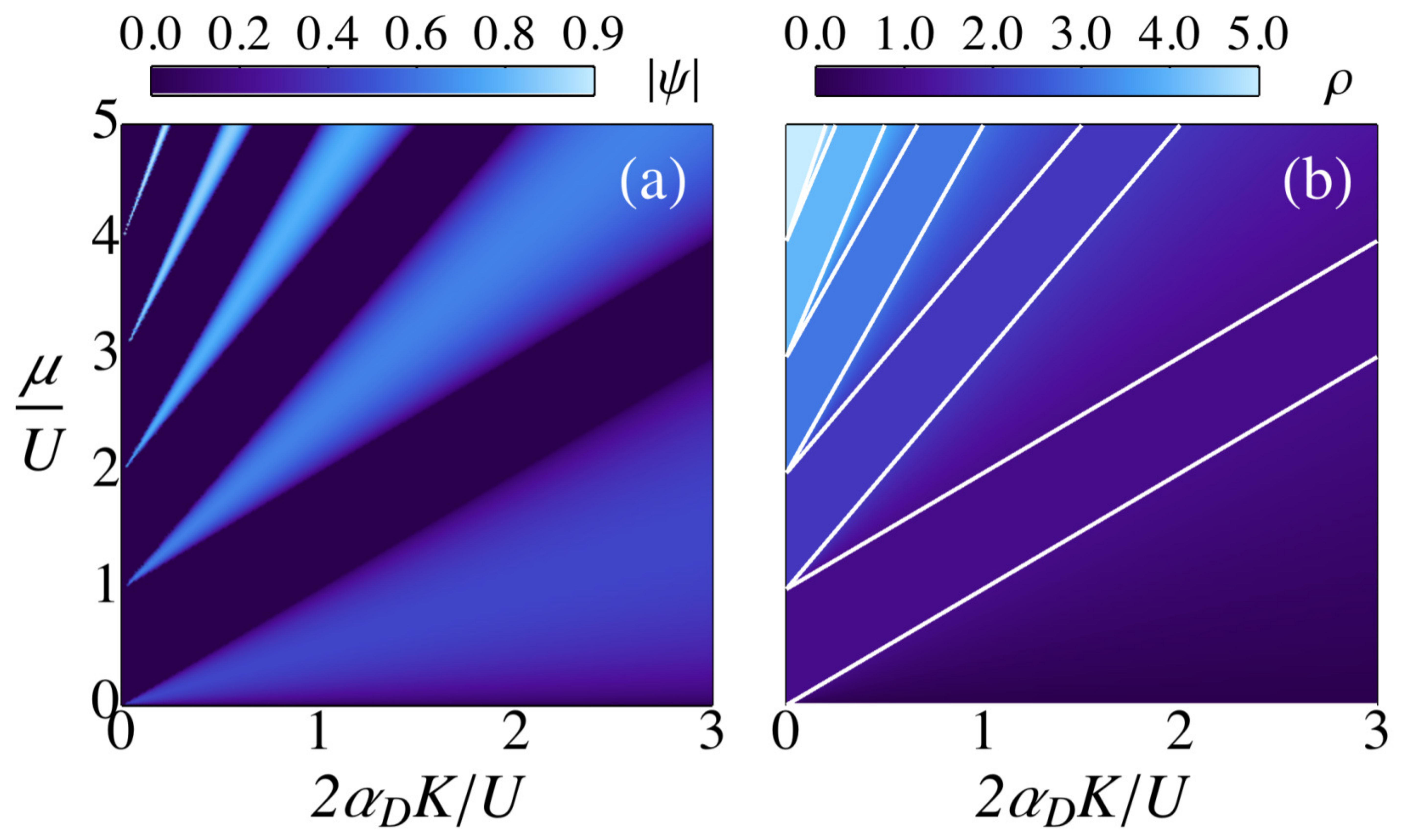}
\caption{ Phase diagram of the bosonic system with quantum optical lattice for homogeneous scattering from the atoms in the limit $t_0/U=0$. (\textbf{a}) Superfluid order parameter: bright areas correspond to the superfluid state with minimal atomic fluctuations and dark areas to Mott-insulator regions. (\textbf{b}) Local density: Mott insulator plateaus surround minimal uncertainty superfluid states with consecutive Fock-state fillings, and white lines denote phase boundaries. }
\label{fig:cones}
\end{figure}

Here, we show that light favours the superfluid state and stabilises it, even in deep optical potentials. Specifically, for a given value of $t_0/U$ corresponding to a Mott insulator state in the absence of quantum light, we find that the ground state of the system becomes a superfluid state with minimal atomic density fluctuations per site, \textit{i.e}.,~only two Fock-state components, with integer fillings $m$ and $m+1$ per site. Therefore, the local atomic density $\uprho=\langle \hat n_i\rangle$ is constrained between these two values: $m<\uprho<m+1$. Considering the case of a very deep optical lattice, where tunnelling can be neglected, the effective on-site energy can be written as:
\begin{equation}\label{energy}
E_{\mathrm{st}}=U (g-1)g/2-Ug x-U\uprho^2/(2\upgamma_{D})
\end{equation}
with $\upgamma_{D}=U/(2 K\upalpha_{D})$, $\upalpha_D= 2g_{\mathrm{eff}}J_D^2$, $g=x+1$, $x=\left(\upmu/U -\uprho/\upgamma_{D}\right)$ and $\upmu$ is the chemical potential.
Expressing Equation (\ref{qpot}) in the Fock state basis, we can diagonalise it and self-consistently derive the on-site density $\uprho=\langle\hat n_i\rangle$, order parameter $\psi=\langle\hat b_i\rangle$ and atomic fluctuations $\Delta(\hat n_i)$ that minimise the ground state energy for a homogeneous system \cite{fisher1989,oosten2001}. The system is a superfluid provided that $\psi\neq0$, and if not, the system is in the Mott insulator state. Specifically, we find that for $m(\upgamma_{D}^{-1}+1)\leq {\upmu}/{U}\leq \upgamma_{D}^{-1}(m+1)+m $, corresponding to the minimal fluctuation superfluid state, these parameters are given~by:
\begin{align}
\psi&=\sqrt{(m+1)(\uprho-m)(1-\uprho+m)} \\
\uprho&= \upgamma_{D}\left(\frac{\upmu}{U} -m\right) \\
\Delta(\hat n_i)&=(\uprho-m)(1-\uprho+m)
\end{align}
while for $ \upgamma_{D}^{-1}(m+1)+m\leq {\upmu}/{U}\leq (m+1)(\upgamma_{D}^{-1}+1) $, one has $\psi=0$, $\uprho=m+1$ and $\Delta(\hat n_i)=0$.
The presence of such a state can be easily seen in the mean occupation $\uprho$, where in between the Mott insulator plateaus, it behaves linearly as a function of the chemical potential. The slope of this function depends on the coupling and, in the large $K$ limit, is $\upgamma_D$. The phase diagram in terms of the effective light strength $\upalpha_D$ shows that Mott insulator states are indeed strongly suppressed (see Figure~\ref{fig:cones}), and in between the Mott insulator plateaus, these minimal fluctuation superfluids emerge.
This behaviour is confirmed by exact diagonalisation simulations where, in the limit of large numbers of illuminated sites, the same pattern is evident. Moreover, we find that $\uprho(\upmu/U)$ has discrete steps depending on the number of illuminated sites $K$ and that they become smooth as $K\to \infty$. This computation can easily be implemented by noting that since we are focussing on the limit $t_0/U=0$ and in absence of a trapping potential (besides the classical optical lattice), the number of basis states needed to find the ground state reduces significantly. Thus, one can easily achieve simulations with 100 or even 1000 sites, and at this point, the continuum limit is evident. We use the symmetries of the effective light-induced long-range interaction in a truncated Hilbert space where the maximum filling per site is five particles; as for the range of the on-site interactions and chemical potential considered, this is sufficient.

 \section{Conclusions}

In summary, we have shown how to use light scattering from ultracold atoms to perform QND measurements on fermionic systems and demonstrated how the quantum addition to the classical diffraction pattern carries information about quantum correlations of the atomic state. Focussing on a single experimental run, we discussed the entanglement properties between light and matter and showed how these can be exploited for creating macroscopic quantum superpositions with fermionic and bosonic systems. We provided different schemes for the realisation of such states and suggested how to make them more robust to decoherence by using a homodyne measurement. Finally, we described the phase diagram of a system of bosons in an optical cavity trapped in an optical lattice. The cavity mediates an effective long-range interaction that stabilises a superfluid state with minimal fluctuations and suppresses the Mott insulator states.


\acknowledgments{Acknowledgements}

The work is financially supported by the EPSRC
 (DTA
 and EP/I004394/1). The authors thank the Jaksch group for use of the TNT library for simulations \cite{TNT}.


\authorcontributions{Author Contributions}

{Thomas Elliott,  Gabriel Mazzucchi, Wojciech Kozlowski and  Santiago Caballero-Benitez} carried out the above work under the supervision of {Igor Mekhov}. All authors contributed to the writing of the~manuscript.


\conflictofinterests{Conflicts of Interest}

The authors declare no conflict of interest.


\end{document}